\pdfoutput=1
\documentclass[]{spie}  %>>> use for US letter paper
\usepackage{amsmath,amsfonts,amssymb}
\usepackage{graphicx}
\usepackage[colorlinks=true, allcolors=blue]{hyperref}

\title{Towards a multi-input astrophotonic AWG spectrograph}

\author[a]{Pradip Gatkine} %  Pradip Gatkine, Yiwen Hu, Tiecheng Zhu, Yang Meng, Sylvain Veilleux, Joss Bland-Hawthorn, Mario Dagenais 
\author[a,b]{Sylvain Veilleux}
\author[c]{Yiwen Hu}
\author[d]{Joss Bland-Hawthorn}
\author[c]{Mario Dagenais}

\affil[a]{Department of Astronomy, University of Maryland, College Park, Maryland 20742, USA}
\affil[b]{Joint Space-Science Institute, University of Maryland, College Park, Maryland 20742, USA}
\affil[c]{Department of Electrical and Computer Engineering, University of Maryland, College Park, Maryland 20742, USA }
\affil[d]{Sydney Institute for Astronomy and Sydney Astrophotonic Instrumentation Labs, School of Physics, The University of Sydney, New South Wales 2006, Australia}

\authorinfo{Further author information: (Send correspondence to P. Gatkine)\\E-mail: pgatkine@astro.umd.edu}

\begin{document} 
\maketitle

\begin{abstract}
Astrophotonics is the new frontier technology to make suitable diffraction-limited
spectrographs for the next generation of large telescopes. Astrophotonic spectrographs are
miniaturized, robust and cost-effective. For various astronomical studies, such as probing
the early universe, observing in near infrared (NIR) is crucial. Therefore, our research
group is developing moderate resolution (R $\sim$ 1500) on-chip photonic spectrographs in the
NIR bands (J Band: 1.1$-$1.4 $\mu m$; H band: 1.45$-$1.7 $\mu m$). To achieve this, we use the concept of arrayed waveguide gratings (AWGs). We fabricate the device using a silica-on-silicon substrate. The waveguides on this AWG are 2 $\mu m$ wide and 0.1 $\mu m$ high Si$_{3}$N$_{4}$ core buried inside a 15 $\mu m$ thick SiO$_{2}$ cladding.

To make the maximal use of astrophotonic integration such as coupling the AWGs with multiple single-mode fibers coming from photonic lanterns or fiber Bragg gratings (FBGs), we require a multi-input AWG design. In a multi-input AWG, the output spectrum due to each individual input channel overlaps to produce a combined spectrum from all inputs. This on-chip combination of light effectively improves the signal-to-noise ratio as compared to spreading the photons to several AWGs with single inputs. In this paper, we present the design and simulation results of an AWG in the H band with three input waveguides (channels). The resolving power of individual input channels is $\sim$1500, while the overall resolving power with three inputs together is $\sim$500, 600, 750 in three different configurations simulated here. The device footprint is only 16 mm $\times$ 7 mm. The free spectral range of the device is $\sim$9.5 nm around a central wavelength of 1600 nm. For the standard multi-input AWG, the relative shift between the output spectra due to adjacent input channels is about 1.6 nm, which roughly equals one spectral channel spacing. In this paper, we discuss ways to increase the resolving power and the number of inputs without compromising the free spectral range or throughput.
\end{abstract}


\section{INTRODUCTION}
\label{sec:intro}  % \label{} allows reference to this section
The field of photonics emerged as a revolution in telecommunication industry in the last two decades. This technology has since found widespread applications in fundamental and applied sciences. The photonic approach is geared to solve to some of the major challenges in observational astronomy \cite{bland2009astrophotonics, bland2017astrophotonics, ellis2017astrophotonics}. As larger and larger ground- and space-based telescopes are being proposed and built, the volume, mass, and cost of the associated conventional optical instruments scale roughly as the cube of the diameter of the telescope \cite{bland2009astrophotonics}. Maintaining the mechanical and thermal stability over such large volumes and across different optical elements of an instrument is challenging. For space telescopes, it is particularly crucial to keep the payload size and weight minimal. The astrophotonic devices are miniaturized and hence, reduce the size and mass of astronomical instrumentation by several orders of magnitude while keeping  the costs substantially lower \cite{bland2017astrophotonics, blind2017spectrographs, bland2006instruments}.\\ 

The power of guiding the light and manipulating it on a micro/nano scale in photonics has imparted immense flexibility to a wide variety of astronomical instruments \cite{ellis2017astrophotonics,bland2017astrophotonics}. These instruments include spectrographs \cite{cvetojevic2012first,cvetojevic2010miniature,gatkine2017arrayed,gatkine2016development}, spatial/spectral filters \cite{huss2005spatial} , Bragg gratings \cite{zhu2016arbitrary,hu2018efficient}, precision wavelength calibrators \cite{murphy2007high}, interferometers \cite{labadie2016astronomical}, nullers \cite{jovanovic2012starlight}, beam combiners \, etc. Along with these instruments, the technology of coupling the inherently multi-mode astronomical light into these primarily single-mode photonic devices has taken a quantum leap. Some of the prominent light-carrying and -coupling devices/technologies include: photonic lanterns \cite{birks2015photonic,leon2010photonic}, pupil remappers \cite{jovanovic2012starlight}, fiber hexabundles \cite{bland2011hexabundles} and multi-core fibers \cite{lindley2014demonstration}. This development is further augmented by the improvements in the adaptive optics (eg. SCExAO \cite{martinache2009subaru}), especially in the near-infrared, enabling a reduction in the number of modes required to efficiently collect the starlight into a multi-/few-mode fiber \cite{horton2007coupling}. All of these recent developments described above are ushering the field of astronomical instrumentation to the photonic era.\\

\subsection{Scientific Motivation}
The photonic spectrograph described here is designed for a specific scientific exploration. Studying the galaxy evolution and the composition of galaxies in the first few billion years of the universe is a unsolved question, which demands deep observations using large area telescopes and efficiently dispersing the collected light to study the signatures of various atomic and ionic species that reveal the composition, physical conditions and kinematics of the material in, around, and between galaxies \cite{gnedin1997reionization,robertson2010early,tanvir2012star,gatkine2018cgm}. Most of these signatures are in the rest-frame ultraviolet (UV) wavelengths. Due to accelerating expansion of the universe, this UV light from the first few billion years of the universe gets redshifted to near-infrared (NIR) light in the observed frame. It is therefore important to focus the investigation of early universe in J and H bands (1150-1350 nm and 1450-1650 nm respectively). The resolving power needs to be high enough to enable meaningful scientific investigation but at the same time it is necessary to maintain a high signal-to-noise ratio, especially since these distant objects tend to be faint. Therefore a moderate resolution ($\lambda$/$\delta\lambda\sim$1500) is required. This kind of exploration requires large telescopes ($>$8 m aperture) and the volume of associated conventional spectrographs scales with the telescope aperture in a superlinear fashion \cite{bland2009astrophotonics,bland2017astrophotonics}. These large-scale instruments face major challenges in mechanical and thermal stability of the optical components. Therefore, implementation of a compact integrated photonic spectrograph is of particular interest in this case. 

%% Adding AWG-conventional-similarity
   \begin{figure} [ht]
   \begin{center}
   \begin{tabular}{c} %% tabular useful for creating an array of images 
   \includegraphics[height=6cm]{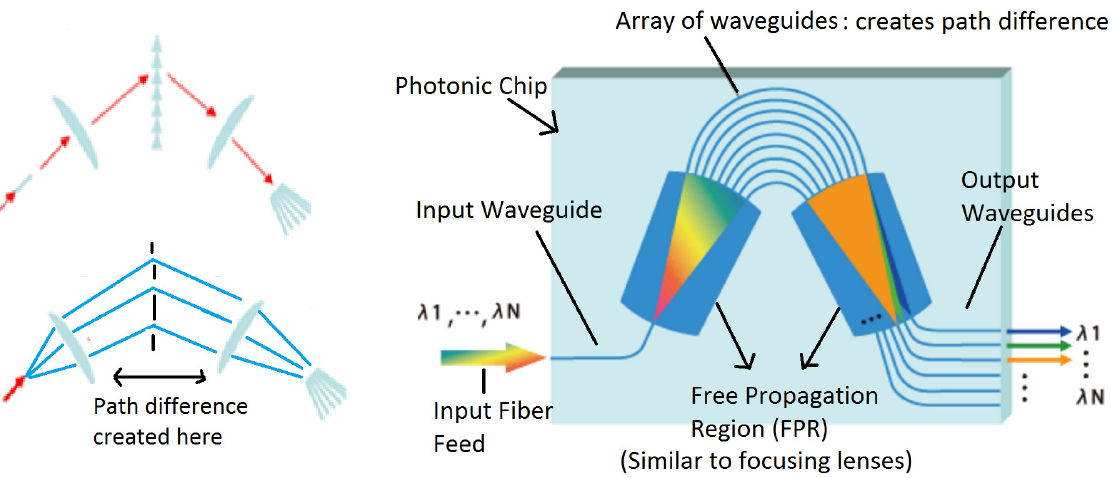}
   \end{tabular}
   \end{center}
   \caption[AWG_sim_conventional] 
%>>>> use \label inside caption to get Fig. number with \ref{}
   { \label{fig:AWG_sim_conventional} 
Analogy between conventional grating spectrograph (left) and arrayed waveguide gratings (right), from Gatkine et al. 2016}
   \end{figure} 

\subsection{Arrayed Waveguide Grating Spectrograph (AWG)}
Among various photonic solutions, arrayed waveguide gratings is a particularly promising approach owing to a well-developed design methodology from telecommunication industry and relative ease of fabrication. An arrayed waveguide grating is, in principle, similar to a conventional diffraction grating spectrograph, as shown in Fig 1 \cite{gatkine2016development}. The input light is guided through the input waveguides to the free propagation region (FPR), which illuminates the array of waveguides with the same phase. The array is constructed such that the path difference between adjacent waveguides is a fixed multiple (i.e. the spectral order) of the wavelength. This light interferes in the output free propagation region leading to constructive interference peaks at the output FPR interface depending on the wavelength. Multiple spectral orders overlap at the output FPR interface depending on the designed free spectral range of the device. The output FPR can either be sampled by output waveguides (as shown in Fig. 1) to measure the spectral composition of the light in specific discrete channels, or the output FPR can be exposed and the overlapping orders can be separated by a low resolution cross disperser (as shown in Fig. 2). This approach gives a more continuous picture of the spectral information, as required in most of the astronomical explorations.  
   
\section{An Integrated Photonic Spectrograph}
The AWG spectrograph described here is the principal component of the integrated astrophotonic spectrograph in the near-IR for the scientific exploration described above using ground-based telescopes. Ground-based near-IR spectroscopy, especially at moderate resolutions is challenging since the near-IR sky gives a bright background from the narrow atmospheric OH-emission lines \cite{ellis2008case, trinh2013gnosis,zhu2016arbitrary}. For a moderate resolution spectrograph, it is important to selectively eliminate these OH-lines prior to dispersion to minimize the noise. We have developed waveguide Bragg gratings (WBGs) to introduce notch filters in the light path to selectively eliminate the OH-emission lines prior to dispersion step in the AWGs \cite{zhu2016arbitrary,hu2018efficient}. These on-chip devices operate in single-mode waveguides, while the telescope illumination is multimode in nature due to atmospheric turbulence. A photonic lantern will be used to efficiently capture and guide the light into several single mode fibers/waveguides \cite{leon2010photonic,spaleniak2014multiband}. The multimode light from the telescope is efficiently coupled into multimode fibers. The photonic lanterns adiabatically taper the multimode fiber into several single-mode fibers (the number depends on quality of the telescope beam). These single-mode fibers will feed the integrated chip with WBGs and AWG on it. Various spectral orders of the dispersed AWG light are separated in the orthogonal direction using a compact low-resolution cross dispersion setup and thus the 2D spectrum will be imaged onto the detector. The schematic of this integrated spectrograph is shown in Fig. 2. This design provides the modularity to stack several of these integrated photonic chips fed by separate lanterns for multi-object spectroscopy or taking spectra of the different wavebands simultaneously using dedicated photonic chips for that particular waveband.

%% Adding AWG-conventional-similarity
   \begin{figure} [ht]
   \begin{center}
   \begin{tabular}{c} %% tabular useful for creating an array of images 
   \includegraphics[height=7cm]{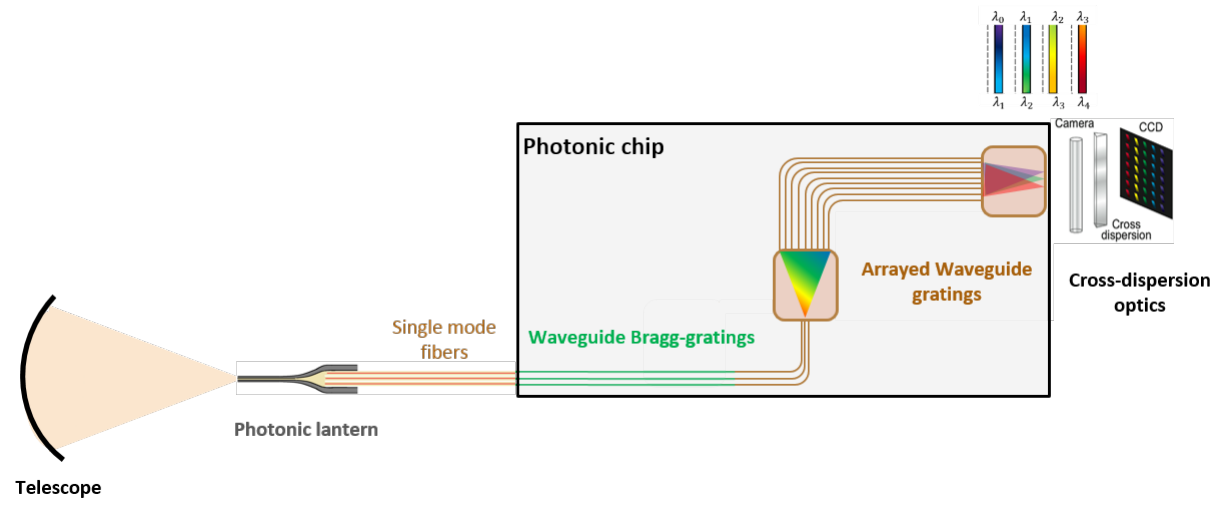}
   \end{tabular}
   \end{center}
   \caption[Full setup of an integrated photonic spectrograph with OH-suppression] 
%>>>> use \label inside caption to get Fig. number with \ref{}
   { \label{fig:Photonic spectrograph} 
A schematic of the full setup of an integrated photonic spectrograph with OH-suppression}
   \end{figure} 
   
   %%%%%%%%% Idea of a figure with nulling interferometer, WBG, and AWG an on the same chip : schematic
   
\subsection{A Few-input AWG}
In Gatkine et al. 2017, we demonstrated a single-input, moderate resolution (R$\sim$1000), high-throughput (peak overall throughput $\sim$ 25\%), broad-band AWG designed for H-band (1450$-$1550 nm). However, capturing the telescope light in a single mode fiber is not highly efficient since the beam is not diffraction limited (unlike the case for radio telescopes) due to atmospheric turbulence \cite{shaklan1988coupling}. Thanks to the recent developments in adaptive optics (eg. SCExAO), it is possible to achieve near-diffraction limited performance, thus allowing efficient coupling of light into a few-mode fiber in near-IR \cite{jovanovic2016efficiently}. In this paper, we are exploring the case of a three-input spectrograph as a test-bench to understand the challenges of a multi-input AWG from an astronomical perspective and propose solutions to effectively combine the light from multiple inputs to achieve a high signal-to-noise ratio spectrum.  

\subsection{Design}
The AWG design is similar to the AWG in Gatkine et al. 2017, but with three input waveguides (instead of one) at the same spatial separation as designed for the output waveguides. This spatial separation is chosen since it is the standard in wavelength division multiplexing application in telecommunication industry where the AWG approach originated \cite{smit1996phasar}. The detailed design, throughput and wavelegnth response of the single-input AWG are described in Gatkine et al. 2017. The physical parameters of the current multi-input AWG are described in Table 1. The design free spectral range of the AWG is 9.5 nm, thus covering the entire H-band in roughly 23 spectral orders. The size of the AWG chip is 16 mm $\times$ 7 mm. 

%% Adding AWG-conventional-similarity
   \begin{figure} [tbp]
   \begin{center}
   \begin{tabular}{c} %% tabular useful for creating an array of images 
   \includegraphics[height=6cm]{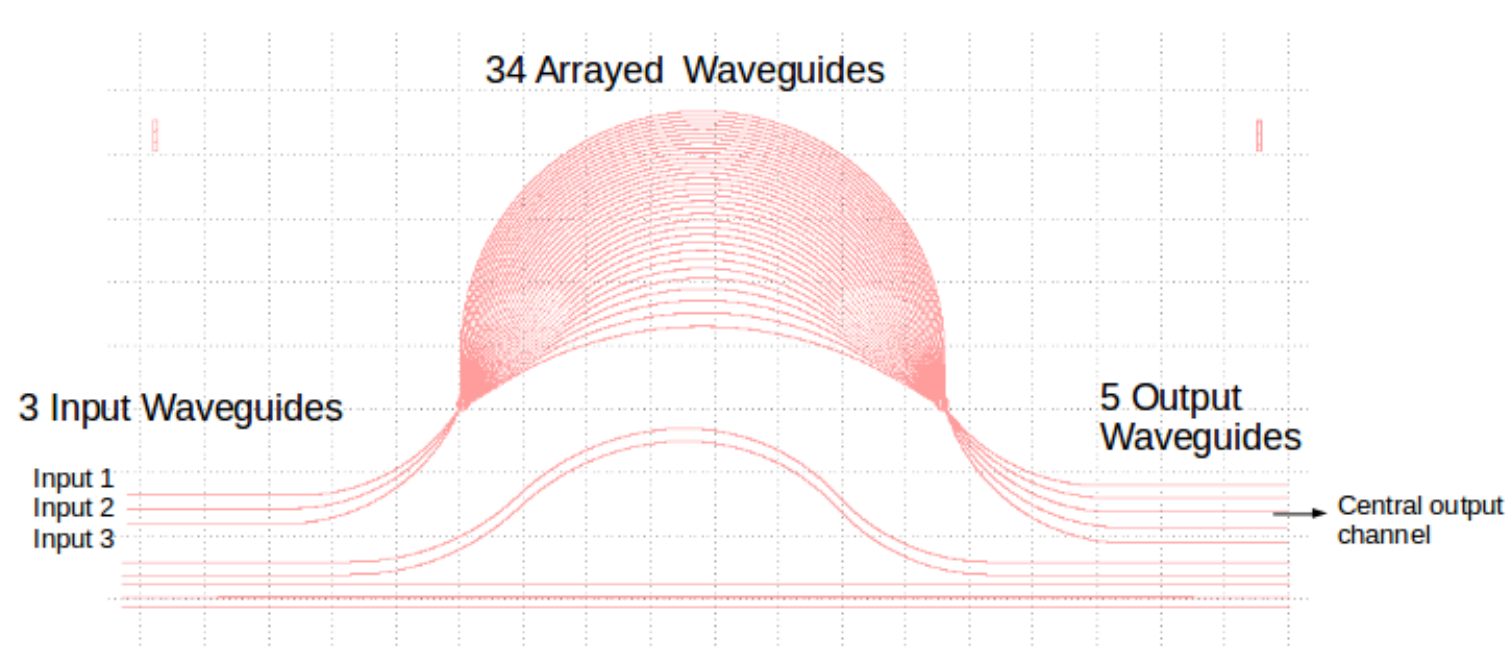}
   \end{tabular}
   \end{center}
   \caption[CAD of the multi-input AWG] 
%>>>> use \label inside caption to get Fig. number with \ref{}
   { \label{fig:AWG CAD} 
The CAD of the designed multi-input AWG. The input waveguides are on the left and the output is sampled by 5 waveguides. There are 34 waveguides in the array. The footprint of the chip is 16 mm $\times$ 7 mm}
   \end{figure} 
   
\begin{table}[tbp]
\caption{Summary of the AWG design} 
\label{tab:AWG design}
\begin{center}       
\begin{tabular}{|l|c|} 
\hline
\rule[-1ex]{0pt}{3.5ex} \textbf{Parameters} & \textbf{Design value} \\
\hline
\rule[-1ex]{0pt}{3.5ex}  1. Spectral resolution for each input ($\lambda/\delta\lambda$) & $\sim$1500 \\
\hline
\rule[-1ex]{0pt}{3.5ex}  2. Free spectral range for each input & $\sim$9.5 nm \\
\hline
\rule[-1ex]{0pt}{3.5ex}  1. Waveguide cross-section & 2.0 $\times$ 0.1 $\mu m$ \\
\hline
\rule[-1ex]{0pt}{3.5ex}  2. Number of waveguides & 34 \\
\hline
\rule[-1ex]{0pt}{3.5ex} 3. FPR length & 200 $\mu m$ \\
\hline
\rule[-1ex]{0pt}{3.5ex} 4. $\Delta L$ & 172 $\mu m$ \\
\hline
\rule[-1ex]{0pt}{3.5ex} 5. Separation between waveguides at array-FPR interface & 6 $\mu m$ \\
\hline
\rule[-1ex]{0pt}{3.5ex} 6. Output waveguide spacing & 6 $\mu m$\\
\hline
\rule[-1ex]{0pt}{3.5ex} 7. Footprint & 16 mm $\times$ 7 mm\\
\hline

\end{tabular}
\end{center}
\end{table}

\subsection{Simulation}
The AWG was simulated using Rsoft software \cite{BeamPROP} to study the characteristics of a representative spectral order centered around 1600 nm. There are 3 input waveguides for this AWG and the output FPR is sampled by 5 output waveguides. In the spectral domain, each of the `discrete' output waveguides are separated by  $\Delta$$\lambda$ $\sim$ 1.6 nm (by design) in all the spectral orders. This is also called `channel spacing’. The AWG is simulated by illuminating each input waveguide one by one and calculating the response at each of the output waveguides. Here, we will focus on the central output channel of the AWG (as shown in Fig.3) and explore how the wavelength response changes with various modifications in the input waveguides. For simplicity, we vary the input waveguide separation progressively as shown in Fig. 4.  

\begin{figure} [tbp]
   \begin{center}
   \begin{tabular}{c} %% tabular useful for creating an array of images 
   \includegraphics[height=10cm]{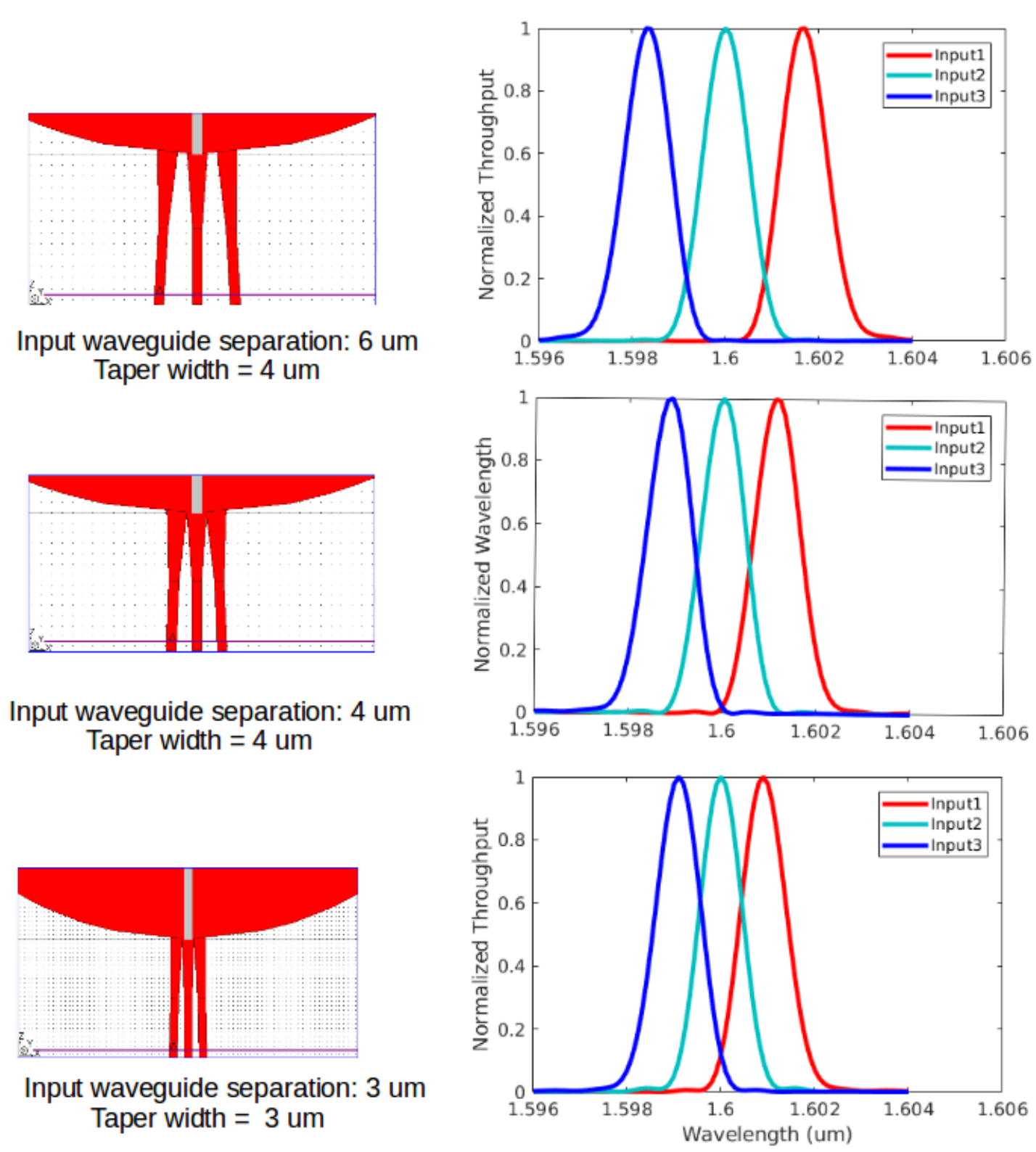}
   \end{tabular}
   \end{center}
   \caption[Simulation results] 
%>>>> use \label inside caption to get Fig. number with \ref{}
   { \label{fig:Simulation_results} 
Simulation results: Each plot shows the normalized throughput of the central output channel in red, green and blue when input 1, 2, and 3 are illuminated respectively. The three cases show the effect of bringing the input waveguides progressively closer by changing the waveguide separation from 3 times the waveguide width (i.e.6 $\mu m$) to 2 times (4 $\mu m$) and further to 1.5 times (3 $\mu m$).}
   \end{figure} 

\subsection{Results}
Each off-center input gives an output spectrum that is shifted in wavelength by an amount (say, $d\lambda$) proportional to the offset from the central input waveguide. This $d\lambda$ shift leads to degradation of the resolving power roughly by a factor of number of input waveguides $\times$ $d\lambda$/$\Delta\lambda$, where $\Delta\lambda$ is the channel spacing as described in the above paragraph (in this case, $\sim$1.6 nm). The goal is to keep this shift minimal to reduce the degradation from the design resolving power. This way, the desired resolution can be achieved without increasing the intrinsic resolution of the AWG by a large factor (to compensate for the degradation). 

Towards that objective, we employ the idea of minimizing the separation between the input waveguides to the maximum possible extent. We demonstrate this idea from the simulation described in Fig. 4. The waveguide separation and the effect on the resolving power degradation is summarized in Fig. 4 and Table 2. It is clear from these simulations that feeding the input light to the AWGs using a compact assembly of waveguides is a very effective approach for minimizing the central wavelength shift and thereby obtaining the desired resolution with minimal need of compensation.   

We are currently developing a photonic spectrograph with a compact input feed to incorporate a few single mode fibers emanating from the photonic lantern attached to an AO-corrected telescope's focus. This demonstration will work as a stepping stone for fabrication and  physically implementation of a higher resolution AWG (by increasing the spectral order) whose final resolution (including the resolution degradation) will be $\sim$1500.

\begin{table}[ht]
\caption{Summary of the AWG simulations} 
\label{tab:AWG design}
\begin{center}       
\begin{tabular}{|c|c|} 
\hline
\rule[-1ex]{0pt}{3.5ex} \textbf{AWG case} & \textbf{Resolving power degradation factor} \\
\hline
\rule[-1ex]{0pt}{3.5ex}  1. Input waveguide separation: 6$\mu m$, taper width: 4$\mu m$ & 1/3 \\
\hline
\rule[-1ex]{0pt}{3.5ex}  2. Input waveguide separation: 4$\mu m$, taper width: 4$\mu m$ & 1/2.5 \\
\hline
\rule[-1ex]{0pt}{3.5ex}  3. Input waveguide separation: 3$\mu m$, taper width: 3$\mu m$ & 1/2 \\
\hline
\end{tabular}
\end{center}
\end{table}

\acknowledgments % equivalent to \section*{ACKNOWLEDGMENTS} 

The authors thank the University of Maryland NanoCenter for the fabrication expertise and the astrophotonics group at University of Sydney and Maquarie University for providing a comprehensive overview of new developments in photonics. The authors thank Prof. Stuart Vogel for his suggestions. The authors acknowledge the financial support for this project from the W. M. Keck Foundation, National Science Foundation, and NASA.

% References
\bibliography{report} % bibliography data in report.bib
\bibliographystyle{spiebib} % makes bibtex use spiebib.bst

\end{document}